\documentclass[
    ,final            
  ]
  {aipproc}

\layoutstyle{8x11double}

\begin{document}

\newcommand{\cV}{\mathcal{V}}
\newcommand{\cP}{\mathcal{P}}
\newcommand{\cH}{\mathcal{H}}
\newcommand{\beware}{\marginpar{\bf beware}}
\newcommand{\beqa}{\begin{eqnarray}}
\newcommand{\eeqa}{\end{eqnarray}}
\newcommand{\beq}{\begin{equation}}
\newcommand{\eeq}{\end{equation}}

\title{Scaling, decoupling and transversality of the gluon propagator}



\author{Christian S. Fischer}{
 address={Institute for Theoretical Physics, 
 University of Giessen, 
 Heinrich-Buff-Ring 1, 35392 Giessen, Germany}
}
\author{Lorenz von Smekal}{
 address={
 Institut for Nuclear Physics, 
 Technische Universit\"at Darmstadt,
 Schlossgartenstra{\ss}e 9, 64289 Darmstadt}
}

\begin{abstract}
In this note we discuss a couple of technical issues relevant to 
solving the Dyson-Schwinger equation for the gluon propagator in
Landau gauge Yang-Mills theory. In the deep infrared functional 
methods extract a one-parameter family of solutions generically
showing a massive behavior referred to as 'decoupling' but also
including the so-called 'scaling' solution with a conformal infrared
behavior as a limiting case. We emphasize that the latter cannot be
ruled out by technical arguments related to the removal of quadratic
divergencies and transversality.
\end{abstract}
\maketitle

\subsection{Introduction}

The infrared behavior of the Green's functions of Landau gauge QCD has
been focus of a number of discussions over the last years. 
In general, this interest is motivated by the desire to understand global
properties of the theory such as the realization of the global gauge
charges which is closely related to the description of confinement, as
formulated by Kugo and Ojima many years ago \cite{Kugo:1979gm},
within the local field theory framework of covariantly gauge-fixed
Yang-Mills theory. Clearly, nonperturbative approaches such as lattice
QCD or functional methods are necessary to test these ideas.

From the functional methods, be it Dyson-Schwinger equations (DSEs) or
the Functional Renormalization Group (FRG), it is now known that two types of
solutions exist which differ in their asymptotic infrared behavior. 
One is a 'massive' or 'decoupling' type of solution, which is
characterized by an infrared finite gluon propagator 
and a ghost propagator with an infrared finite dressing function 
\cite{Aguilar:2008xm,Boucaud:2008ji,Dudal:2008sp,Fischer:2008uz}, and
the other is the so-called 'scaling' solution with unique infrared power
laws for both propagators 
\cite{von Smekal:1997vx,Lerche:2002ep,Zwanziger:2002ia,Pawlowski:2003hq}
and all other Green's functions of Landau gauge Yang-Mills theory
\cite{Alkofer:2004it,Fischer:2006vf}.

In terms of the dressing functions $G(p^2)$ 
and $Z(p^2)$ of the ghost and gluon propagators in Landau gauge 
\begin{equation}
D_G(p) = -\frac{G(p^2)}{p^2}\,, \hspace*{1mm} 
 D_{\mu \nu}(p) =  \left(\delta_{\mu \nu} -
    \frac{p_\mu p_\nu}{p^2}\right)
  \frac{Z(p^2)}{p^2}\,,
 \label{props} 
\end{equation}
decoupling is characterized by the infrared behavior
\begin{equation}
Z(p^2) \sim p^2/M^2 \; ,  \hspace*{.2cm} 
G(p^2) \to \mathrm{const.} \; , \;\; \mbox{for} \;\; p^2 \to 0,  
\label{typeII} 
\end{equation} 
whereas with scaling in 4 dimensions one has, 
\begin{equation}
Z(p^2) \sim (p^2)^{2\kappa} \; , \hspace*{.2cm} 
G(p^2) \sim (p^2)^{-\kappa} \; ,  \label{typeI} 
\end{equation} 
with a positive exponent $\kappa < 1$ which under a certain regularity
assumption on the ghost-gluon vertex \cite{Lerche:2002ep} results to
be $\kappa = \kappa_c = (93 - \sqrt{1201})/98 \approx 0.6$.

Both, scaling and the decoupling type of solutions together form a 
one-parameter family, which has been obtained analytically and as 
infrared limits of numerical solutions to functional equations. With 
fixed gluon input this family has been found in the ghost DSE in
Ref.~\cite{Boucaud:2008ji}; full numerical solutions for the coupled
ghost and gluon system of DSEs and FRGEs have been given for the whole
one-parameter family in Ref.~\cite{Fischer:2008uz}. The decoupling
results agree quantitatively well with lattice data \cite{Cucchieri:2008fc}. 
Scaling has not been observed unambiguously on the lattice so far.
There is an ongoing effort, however, to understand why that is the
case and what the differences are between the functional continuum methods
and those commonly used in what is called lattice Landau gauge
\cite{vonSmekal:2008ws,Sternbeck:2008mv,Maas:2009se}.  

Meanwhile, we would like to reply in this note to a claim made by Mike
Pennington concerning an apparent problem with scaling from continuum
arguments alone. In his plenary talk at this conference he purported
that infrared scaling was observed only as a result of an
inconsistency between the way in which quadratic divergences are being
removed and the transversality of the gluon DSE in Landau gauge.  
In the following we explain why this conclusion is itself
incorrect. There is no problem with quadratic divergences and
transversality in the scaling results from the functional continuum
methods.     

\subsection{Transversality of the gluon propagator}

In order to understand the matter let us first recall, that
in covariant gauges there is a Slavnov-Taylor identity for the gluon propagator
stating that its longitudinal part is not modified by interactions,
\beq
-\partial_\mu \partial_\nu D_{\mu \nu}^{ab}(x-y) = \xi \delta^{ab}
\delta^4(x-y) \; . \label{trans}
\eeq
In the Landau gauge limit $\xi \rightarrow 0$ it then follows that the
fully dressed gluon propagator remains transverse which implies that
its momentum space Dyson-Schwinger equation is of the form,
\beq
\cP_{\mu \nu}^T \frac{p^2}{Z(p^2)} = \cP_{\mu \nu}^T p^2 Z_3 + \Pi_{\mu \nu}(p^2) \label{DSE}
\eeq 
where $Z_3$  is the gluon renormalization constant, $\cP_{\mu \nu}^T =
\left(\delta_{\mu\nu} - {p_\mu p_\nu}/{p^2}\right)$ the transverse
projector, and $\Pi_{\mu\nu}(p^2)$ the self-energy  which is then necessarily
transverse also.
In practical calculations, however, the transversality of
$\Pi_{\mu\nu}$ is often difficult to maintain in specific truncations.
The main sources of transversality violations are thereby the following: 
(i) Numerical solutions of dimensionally regularized
integral equations are extremely cumbersome
\cite{Kizilersu:2001pd}. In practice one therefore  
uses different schemes such as momentum subtractions with a hard cutoff.
While these schemes in general preserve multiplicative
renormalizability of the theory, they violate Eq.~(\ref{trans}). (ii) Ansaetze for the different vertices in $\Pi_{\mu\nu}(p^2)$, 
necessary to close the gluon DSE, may be inadequate to preserve transversality.
In both cases, (i) and (ii), artificial longitudinal contributions $\cP^L_{\mu \nu} L(p^2)$ 
arise on the right hand side of the gluon DSE,
\beq
\Pi_{\mu\nu}(p^2) = \cP_{\mu \nu}^T \Pi(p^2) + \cP_{\mu \nu}^L L(p^2) 
\label{gluon}
\eeq 
with $\cP_{\mu \nu}^L = p_\mu p_\nu / p^2$. 
A different but closely related problem is the appearance of quadratic
divergences in $\Pi_{\mu\nu}$. Again, these are absent in
dimensional regularization, but they occur with the cutoff procedure
typically used in numerical studies of the gluon DSE
\cite{Aguilar:2008xm,Fischer:2008uz,Fischer:2002hna}. 

When discussing the non-perturbative infrared properties of the gluon,
care must be taken that artifacts like non-transversality or
the appearance of quadratic divergencies do not affect the
results. This may be particularly important when it comes to the
discussion of scaling vs decoupling, Eqs.(\ref{typeII}),
(\ref{typeI}).  
In his talk, Mike Pennington has addressed this problem, claiming that 
{\it the very appearance of the scaling solution (\ref{typeI}) is such
an artifact}.  His argument relies on the assumption that the ghost-gluon 
vertex is essentially bare in the infrared. If one then uses a particular 
method to remove quadratic divergencies (see Eq.~(\ref{proj}) below)
the scaling solution disappears. 

This problem has been addressed already in Ref.~\cite{Lerche:2002ep}: Scaling goes 
hand-in-hand with the 
infrared dominance of ghosts, i.e. the ghost-loop dominates
$\Pi_{\mu\nu}(p^2)$, {\it i.e.}, for $ p^2 \to 0$. 
Therefore, the ghost-loop must itself be transverse in the infrared. 
In a truncation where the full ghost-gluon vertex is replaced by the
tree-level vertex of standard Faddeev-Popov theory this would require
$\kappa = 3/4$, which is incompatible with the self-consistently obtained value
$\kappa_c \approx 0.6$ mentioned above. Thus the non-perturbative
ghost-gluon vertex can not be identical to the tree-level one.

Instead, infrared dominance of ghosts and with that the scaling solution
essentially requires that the fully dressed ghost-gluon vertex becomes
itself transverse with respect to the gluon momentum in the infrared
\cite{Lerche:2002ep}. Under the mild additional regularity assumption 
one then obtains the self-consistent value $\kappa= \kappa_c \approx 0.6$. 
Other values $1/2 < \kappa < 1$ are possible when this condition is relaxed
\cite{Lerche:2002ep}. Note, that all these analytic results were obtained in
dimensional regularisation. They are thus unaffected by any technical
difficulties that might arise with quadratic divergences in numerical
investigations.

One method to remove quadratic divergences in the gluon DSE goes back to 
Brown and Pennington \cite{Brown:1988bn}. They decomposed the gluon
self-energy into 
\beq
\Pi_{\mu\nu}(p^2) = \delta_{\mu \nu} F_1(p^2) - p_\mu p_\nu
F_2(p^2)\; , \label{BP}
\eeq
and noted that quadratic divergences can only occur in the term proportional to
$\delta_{\mu \nu}$. 
Consequently, they suggested to project out this contribution. This is done 
by contracting the gluon-DSE with a general projector
\cite{Fischer:2002hna} 
\beq
\cP_{\mu \nu}^\zeta = \Big(\delta_{\mu \nu} -\zeta \frac{p_\mu
  p_\nu}{p^2}\Big)\; \label{proj},
\eeq
and setting $\zeta=4$ in $d=4$ dimensions. This removes quadratic divergencies
and leads to a dependence of $Z(p^2)$ on $F_2(p^2)$ alone. 
Alas, let us see what one obtains for general
$\zeta$ upon contracting the gluon-DSE in the form of Eqs.~(\ref{DSE}) and (\ref{gluon}) 
with $ \cP_{\mu \nu}^\zeta$:
\beq 
\frac{p^2}{Z(p^2)} = p^2 Z_3 + \Pi(p^2) + \frac{1-\zeta}{3} L(p^2) \;
. \label{proj2}
\eeq
Clearly, if the gluon DSE is transverse, $L(p^2)=0$, and the resulting 
Eq.~(\ref{proj2}) is independent of the parameter $\zeta$ and
therefore also free from quadratic divergences. Conversely, the required
$\zeta$-independence provides a valuable test of specific truncations
in numerical studies. 

Two transverse truncations passing this test have been
constructed explicitly in the literature, and both of them 
rely on a nontrivial ghost-gluon vertex thus dismissing Mike's
criticism of the inconsistency that arises with the bare one. 
One is based on the Pinch-Technique/Background Field Method, see
\cite{Binosi:2009qm}, for which the truncation
of Ref.~\cite{Aguilar:2008xm} has also been analyzed analytically in 
dimensional regularization.\footnote{Note, however, that the numerical
  solutions presented in Ref.~\cite{Aguilar:2008xm} have been obtained
  using a hard UV-cutoff and therefore need to be checked separately
  for transversality.} The other approach is the one 
used in Ref.~\cite{Fischer:2008uz}. There, Ansaetze for the
ghost-gluon and three-gluon vertices have been constructed explicitly 
such that the gluon-DSE is free from quadratic divergencies and the
residual contributions to $L(p^2)$ are minimized. The ghost-gluon
vertex used there reduces in the infrared to the minimally dressed
transverse one proposed in \cite{Lerche:2002ep}.
It contains a bare part and  a nontrivial longitudinal part that
serves to cancel all contributions to $L(p^2)$, thus explicitly
establishing {\it exact} transversality of the ghost-loop in the infrared. 
Numerically, the remaining contributions to $L(p^2)$ for all momenta
$p^2$ stay well below one percent of $\Pi(p^2)$. In particular $L(p^2) 
\rightarrow 0$ for $p^2 \rightarrow 0$. This therefore establishes
unambiguously that the DSE results of \cite{Fischer:2008uz} are {\it not}
affected by transversality violating artifacts:  
there {\it is} a one-parameter family of solutions in the infrared including 
decoupling {\it and} scaling solutions. 

The earlier truncation scheme from Ref.~\cite{Fischer:2002hna}, which was   
particularly criticized by Mike Pennington in his plenary talk, was
indeed less far developed and not manifestly transverse. The
artificial longitudinal term $L(p^2)$ in  Eq.~(\ref{proj2}) was removed
by transverse projection, using $\zeta=1$ rather than $\zeta=4$ as suggested
by Brown and Pennington before, which meant that quadratic divergences
had to be dealt with in a different manner. Even though this procedure
might seem less elegant, there were good reasons at the time for that
given in \cite{Lerche:2002ep}. Before we briefly explain those, it can
now be verified a posteriori, by comparing the earlier results from
\cite{Fischer:2002hna} to those of Ref.~\cite{Fischer:2008uz}, 
to verify that they are almost identical (for $\zeta=1$). 

The reason
why transverse projection onto $\Pi(p^2)$ in Eq.~(\ref{gluon}) 
must be done \cite{Lerche:2002ep}
rather than $\zeta=4$ projection onto $F_2(q^2)$ in (\ref{BP}), in a not
manifestly transverse truncation where this can make a difference,
relates to an ambiguity in the Landau gauge: the existence of a second
gauge parameter which interpolates between standard Faddeev-Popov
theory and the ghost/anti-ghost symmetric Curci-Ferrari gauges. In
Landau gauge there is no distinction, however, and any quantity
depending on this parameter is thus ambiguous. In
\cite{Lerche:2002ep} it was shown that $F_2(p^2)$ is such a quantity
while $\Pi(p^2)$ is not. So we have understood over the years that 
it is the original proposal from \cite{Brown:1988bn} to avoid quadratic
divergences which is ambiguous, unfortunately. Luckily, however, the
problem is now completely solved with the manifestly transverse
truncations that were developed since then. 

Finally, we emphasize that the problem with quadratic divergences
is completely unheard of in the FRGEs which are finite by
construction. Yet, one still finds the full one-parameter family of
scaling and decoupling solutions
\cite{Pawlowski:2003hq,Fischer:2008uz}. Apart from the extreme infrared,
where this parameter matters, these solutions all agree and are
furthermore in almost perfect agreement with lattice data.

\subsection{Conclusions}
In this note we have pointed out that a claim made by Mike Pennington in his
plenary talk is not correct. He purported that the infrared scaling solution
for the ghost and gluon propagators of Landau gauge Yang-Mills theory is a mere 
artifact of a technical inaccuracy in the treatment of the gluon
DSE. This is based an the overly simplistic truncation which cannot
be combined with an earlier proposal to remove artificial quadratic
divergences in an untlraviolet cutoff regularization. This issue has
been completely solved over the years, however. The infrared-scaling
solution for the Yang-Mills sector of QCD does not have fundamental
problems related to quaderatic divergences or transversality. 

\vspace{.2cm}

\noindent {\bf Acknowledgements:} 
This work has been supported by the Helmholtz Young Investigator
Grant VH-NG-332 and the Helmholtz International Center for FAIR 
within the LOEWE program of the State of Hesse. 

\vspace*{-.2cm}

\bibliographystyle{aipproc}   

\end{document}